\documentclass[aps,nofootinbib,prd]{revtex4-2}

\usepackage{epsfig,epsf}
\usepackage{amsmath}
\usepackage{amsthm}
\usepackage{amsfonts}
\usepackage{amssymb}
\usepackage{dsfont}
\usepackage{multirow}
\usepackage{appendix}
\usepackage{slashed}
\usepackage[active]{srcltx}
\usepackage{psfrag}
\usepackage{subfigure}
\usepackage[colorlinks,citecolor=blue,urlcolor=red,linkcolor=red]{hyperref}
\usepackage{tabularx}
\usepackage{booktabs}
\usepackage{array}
\usepackage{siunitx}
\usepackage{xcolor}
\usepackage{geometry}
\usepackage{microtype}

\begin{document}

\title{Heavy-quark limit of $B$-meson distribution amplitudes in light-front holography with longitudinal confinement}

\author{R. Khosravi}
\affiliation{Department of Physics, School of Science, Shiraz
University, Shiraz 71946-84795, Iran}

\email{rezakhosravi@shirazu.ac.ir}

\begin{abstract}
The heavy-quark limit of light-front holographic wave functions for
heavy--light mesons is investigated, with particular emphasis on the
leading-twist $B$-meson light-cone distribution amplitude,
$\phi_B^+(\omega)$. Two prescriptions for incorporating finite quark
masses are compared: the invariant mass ansatz (IMA) and the
longitudinal confining framework based on the Li--Maris--Zhao--Vary
(LMZV) potential. In the IMA, although the peak of the distribution
amplitude remains at a soft spectator momentum, its characteristic
width decreases as $\Delta\omega \propto m_B^{-1/2}$ in the
heavy-quark limit, causing the longitudinal distribution to collapse
asymptotically to $\delta(\omega-m_q)$. Accordingly, the inverse
moment approaches the finite limit $\lambda_{B,\mathrm{IMA}} \to m_q
= \mathcal{O}(\Lambda_{\mathrm{QCD}})$. In contrast, the LMZV
framework approaches a finite leading-power (LP) profile with a
stable width of order $\mathcal{O}(\Lambda_{\rm QCD})$ and a stable
inverse moment, $\lambda_{B,\mathrm{LMZV}}^{\mathrm{LP}} = m_q +
\sigma/2 = \mathcal{O}(\Lambda_{\mathrm{QCD}})$, which is
independent of the heavy-quark mass. The heavy-quark scaling of
$f_B$ is also examined using the local axial-vector current at zero
transverse separation. While the LMZV longitudinal wave function has
a non-singular heavy-quark limit, the bare light-front current does
not reproduce the canonical heavy-quark effective theory scaling
$f_B \propto m_B^{-1/2}$: the decay constant scales as $f_B \propto
m_B^{-1}$ in the LMZV framework and as $f_B \propto m_B^{-5/4}$ in
the IMA. Thus, a non-singular heavy-quark limit of the longitudinal
wave function is necessary but insufficient to ensure the correct
heavy-quark scaling of local current matrix elements without
systematic current matching.
\end{abstract}

\maketitle

\section{Introduction}
\label{sec1}

Exclusive decays of heavy $B$-mesons provide a suitable testing
ground for the Standard Model (SM) of particle physics, enabling
precise determinations of the Cabibbo--Kobayashi--Maskawa (CKM)
matrix elements and serving as sensitive probes of potential new
physics beyond the SM \cite{Buchalla1995,Artuso2009,Brod2014}. In
the framework of QCD factorization (QCDF)
\cite{Beneke1999,Beneke2000} and Soft-Collinear Effective Theory
(SCET) \cite{Bauer2000,Bauer2001,Beneke2002}, hard exclusive
processes involving heavy--light mesons, such as the radiative
leptonic decay $B \to \gamma \ell \nu$ and the semileptonic
transition $B \to \pi \ell \nu$, can be systematically factorized
into perturbatively calculable hard-scattering kernels convoluted
with universal, nonperturbative light-cone distribution amplitudes
(LCDAs) \cite{Grozin1996,Beneke2000,DescotesGenon2002}.

Within this collinear factorization framework, the LP contribution
to these amplitudes is governed by the leading-twist $B$-meson LCDA,
$\phi_B^+(\omega)$, where we use $\omega=xm_B$ as the light-front
momentum variable carried by the soft spectator quark. In
particular, a key hadronic uncertainty in radiative leptonic $B$
decays enters through the first inverse moment, $\lambda_B^{-1}$,
which provides a fundamental nonperturbative input to the
hard-collinear factorization theorems
\cite{Korchemsky1999,Lunghi2002,Bosch2003,Beneke2011,Wang2018}.
Owing to the $1/\omega$ weighting, $\lambda_B$ is particularly
sensitive to the soft endpoint behavior of the distribution as
$\omega\to 0$. Despite extensive efforts using QCD sum rules
\cite{Braun2003,Khodjamirian2006,Gubernari2018} and lattice-QCD
developments \cite{Kawamura2018,Wang2019}, a first-principles
determination of $\phi_B^+(\omega)$ and a fully controlled
nonperturbative determination of $\lambda_B$ remain challenging.
Consequently, relativistic bound-state approaches capable of
providing a self-consistent description of heavy--light mesons and
their light-front structure remain of considerable interest.

Light-front holographic QCD (LFHQCD)
\cite{Brodsky2006,DeTeramond2008,Brodsky2014} provides a useful
nonperturbative framework in which hadronic structure is described
in terms of light-front wave functions (LFWFs), with confinement
incorporated through an effective potential. For light mesons
composed of approximately equal-mass constituents, the introduction
of a quadratic dilaton profile reproduces linear Regge trajectories
and provides a successful description of transverse confinement
dynamics \cite{Karch2006,Brodsky2008,Brodsky2008a}. Extending this
framework to heavy--light bound states such as the $B$-meson,
however, requires particular care because of the pronounced
hierarchy between the bottom-quark mass, $m_b$, and the light
spectator-quark mass, $m_q$. In the heavy-quark regime, the
spectator momentum fraction is parametrically localized in the soft
endpoint region, $x\sim\Lambda_{\mathrm{QCD}}/m_B\ll 1$, making the
structure of the longitudinal dynamics essential for establishing a
consistent heavy-quark limit.

In the standard holographic literature, the heavy--light mass
asymmetry is commonly incorporated through the IMA
\cite{Brodsky2014,Ahmady2018,Ahmady2020}, in which the massless
light-front kinetic term is replaced by the off-shell invariant mass
of the valence constituents. Although the IMA successfully generates
an endpoint-concentrated spectator distribution, its asymptotic
behavior in the heavy-quark limit, $m_b \to \infty$, requires closer
examination \cite{Isgur1989,Neubert1994}. A consistent heavy-quark
description involves two distinct conditions: first, it should
maintain a finite soft longitudinal scale for the light degrees of
freedom---ensuring both a stable inverse moment $\lambda_B \sim
\Lambda_{\mathrm{QCD}}$ and a well-defined profile width in
$\omega$. Second, the physical local current matrix element should
recover the heavy-quark effective theory (HQET) scaling $f_B
\sqrt{m_B} = \mathrm{const.}$, with the appropriate short-distance
matching taken into account. As these conditions address
fundamentally different aspects of the heavy-quark limit, they
require independent verification.

We conduct a systematic investigation of these properties by
comparing the standard IMA prescription with a dynamical
longitudinal variational construction based on the LMZV effective
confining potential \cite{Li2016}. In the IMA, the position of the
distribution peak remains tied to the soft spectator scale, with
$\omega_0 \to m_q$ as $m_b \to \infty$. However, for a fixed
transverse confinement scale $\kappa$, the characteristic momentum
width decreases as $\Delta\omega \propto m_B^{-1/2}$, causing the
longitudinal distribution to become increasingly narrow and approach
the singular limiting form $\delta(\omega - m_q)$. Nevertheless, the
inverse moment remains finite, with $\lambda_{B,\mathrm{IMA}} \to
m_q = \mathcal{O}(\Lambda_{\mathrm{QCD}})$, demonstrating that the
essential shortcoming of the IMA is the collapse of the longitudinal
width rather than the vanishing of the inverse moment.

In contrast to the IMA, the LMZV longitudinal construction provides
an alternative realization of the heavy--light dynamics in which the
longitudinal confinement scale $\sigma$ enters explicitly as an
independent parameter \cite{Li2016}. The ground-state mode
consequently approaches a finite LP profile whose characteristic
width is set by $\sigma\sim\mathcal{O}(\Lambda_{\rm QCD})$. In this
scheme the distribution peak is shifted to $\omega_0 \to m_q +
\sigma/2$, fixing the inverse moment at the finite heavy-quark limit
$\lambda_{B,\mathrm{LMZV}}^{\mathrm{LP}} = m_q + \frac{\sigma}{2} =
\mathcal{O}(\Lambda_{\rm QCD})$. With this construction the
longitudinal width does not collapse as $m_b\to\infty$: the profile
retains a physical, $\Lambda_{\rm QCD}$-size spread rather than
degenerating into $\delta(\omega - m_q)$. The two prescriptions
therefore lead to parametrically different longitudinal limits, even
though both keep a soft spectator momentum scale; this distinction
provides a direct test of whether the longitudinal dynamics remain
stable as the heavy-quark mass is increased.

The heavy-quark behavior of the decay constant provides a further
and independent test of the same wave functions. Within the minimal
light-front current formalism evaluated at zero transverse
separation ($\mathbf{b}_\perp = 0$), the matrix element is governed
by the longitudinal overlap integral $\int_0^1 dx\, X(x)$. Under
this prescription, the decay constant scales as $f_B\propto
m_B^{-5/4}$ in the IMA and as $f_B\propto m_B^{-1}$ in the LMZV
framework. These results allow the longitudinal and
current-matrix-element aspects of the heavy-quark limit to be
distinguished clearly. In particular, a finite and stable LP
longitudinal distribution does not, by itself, ensure the canonical
HQET scaling of a local current matrix element. For the IMA,
reproducing $f_B\sim m_B^{-1/2}$ within the present light-front
setup would require an explicit heavy-quark dependence of the
transverse confinement parameter $\kappa$. In the LMZV framework,
the corresponding mismatch shows that the bare light-front current
alone does not reproduce the HQET scaling, so that a systematic
matching to the appropriate effective-theory current is needed
before a physical heavy-quark-limit prediction for $f_B$ can be
claimed.

The analysis presented in this work therefore aims at examining the
heavy-quark limit of holographic $B$-meson wave functions from these
two complementary perspectives: the longitudinal distribution, on
the one hand, and the current-matrix-element formalism, on the
other. The IMA and LMZV prescriptions are studied at finite
heavy-quark mass and in their respective asymptotic limits, with
particular attention to the quantitative observables---the peak
position, the longitudinal width, the limiting form of the LCDA, and
the inverse moment $\lambda_B$. The same wave functions are then
employed to probe the heavy-quark behavior of the decay constant
within the light-front current formalism. In this way, the
conditions required for a stable, soft longitudinal profile are
distinguished from---rather than conflated with---those required for
the correct scaling of a local current matrix element. The
discussion thereby keeps the focus on the heavy-quark consistency of
the holographic description, without treating the mere existence of
a finite $\lambda_B$ as a sufficient criterion.

The remainder of this paper is organized as follows. In Sec.
\ref{sec2}, the light-front holographic framework and the general
structure of the valence wave functions for heavy--light mesons are
reviewed. In Sec. \ref{sec3}, the leading-twist $B$-meson LCDAs are
constructed and their first inverse moments are evaluated within the
IMA and LMZV longitudinal formalisms. In Sec. \ref{sec4}, the
heavy-quark asymptotic behavior is analyzed, including the
longitudinal width, the limiting LCDA, and the scaling of the decay
constant. Finally, Sec. \ref{sec5} summarizes the main conclusions
and discusses possible directions for further development.

\section{Light-front holographic framework}
\label{sec2}

At the level of an effective light-front Hamiltonian, the
bound-state equation for a quark--antiquark meson can be written
schematically as
\begin{equation}
\left[ -\frac{\nabla_{\mathbf{b}_\perp}^{2}}{x(1-x)}
+\frac{m_q^2}{x} +\frac{m_{\bar{q}'}^2}{1-x} +U(x,\mathbf{b}_\perp)
\right]\Psi(x,\mathbf{b}_\perp)=M^2\Psi(x,\mathbf{b}_\perp),
\label{eq201}
\end{equation}
where $M$ is the physical meson mass, $m_q$ and $m_{\bar{q}'}$ are
the constituent quark masses, and $U(x,\mathbf{b}_\perp)$ denotes an
effective interaction that contains the confining dynamics. For the
heavy--light $B$-meson system analyzed in this work, the heavy quark
is identified as the bottom quark with mass $m_b$, while $m_q$
represents the light spectator quark carrying the longitudinal
momentum fraction $x$ (with the $b$ quark carrying $1-x$), separated
by the transverse coordinate $\mathbf{b}_\perp$.

LFHQCD provides a semiclassical framework for relating this dynamics
to an effective description in a higher-dimensional AdS background
\cite{Brodsky2006,DeTeramond2008}. For finite quark masses and a
generic longitudinal interaction, Eq. (\ref{eq201}) is not exactly
separable into independent transverse and longitudinal equations.
The factorized representation introduced below is therefore
understood as a semiclassical holographic approximation, where the
boost-invariant transverse variable
$\zeta=\sqrt{x(1-x)}\,|\mathbf{b}_\perp|$ is identified with the AdS
fifth-dimensional coordinate ($z$) and captures the dominant
transverse dynamics \cite{DeTeramond2008,Brodsky2014}.

Following the standard light-front holographic construction
\cite{Brodsky2008,DeTeramond2008}, the valence LFWF is written in
factorized form as
\begin{equation}
\Psi(x,\zeta,\phi_b) =\frac{e^{iL\phi_b}}{\sqrt{2\pi\zeta}}
X(x)\phi(\zeta), \label{eq202}
\end{equation}
where $L$ is the internal orbital angular momentum, $\phi_b$ is the
azimuthal angle in transverse coordinate space, $X(x)$ describes the
longitudinal dynamics, and $\phi(\zeta)$ contains the transverse
dynamics. In the massless-quark semiclassical limit, and after
separating the longitudinal and transverse variables, the transverse
dynamics is described by a one-dimensional Schr\"odinger-like
equation,
\begin{equation}
\left[ -\frac{d^2}{d\zeta^2} -\frac{1-4L^2}{4\zeta^2}
+U_\perp(\zeta) \right]\phi(\zeta) =M_\perp^2\phi(\zeta),
\label{eq203}
\end{equation}
where $M_\perp^2$ denotes the eigenvalue of the transverse
semiclassical holographic problem. For finite quark masses, the
physical meson mass receives additional contributions from the
longitudinal and quark-mass terms and should not, in general, be
identified with $M_\perp$ alone.

In the soft-wall realization of holographic QCD, conformal symmetry
is broken through a quadratic dilaton term $\kappa^2 z^2$
\cite{Karch2006,Brodsky2008}, where $\kappa$ sets the characteristic
confinement scale. The resulting light-front effective potential
takes the harmonic form \cite{Brodsky2008,DeTeramond2008}
\begin{equation}
U_\perp(\zeta) =\kappa^4\zeta^2 + 2\kappa^2(L+S-1), \label{eq204}
\end{equation}
where $S$ is the total quark spin. This potential leads to linear
Regge trajectories and provides a useful semiclassical description
of the transverse dynamics of mesons. For heavy--light systems,
however, the longitudinal mass-dependent dynamics must be
incorporated separately \cite{Brodsky2014}.

The normalizable solutions of Eq. (\ref{eq203}) are
\cite{Brodsky2008}
\begin{equation}
\phi_{n_\perp, L}(\zeta) =\kappa^{1+L} \sqrt{\frac{2
n_\perp!}{(n_\perp+L)!}} \zeta^{1/2+L} e^{-\kappa^2\zeta^2/2}
L_{n_\perp}^L(\kappa^2\zeta^2), \label{eq205}
\end{equation}
with eigenvalues
\begin{equation}
M_\perp^2(n_\perp,L,S) =4\kappa^2 \left( n_\perp+L+\frac{S}{2}
\right), \label{eq206}
\end{equation}
where $L_{n_\perp}^L$ denotes an associated Laguerre polynomial and
$n_\perp$ is the radial (transverse) quantum number. For the
ground-state valence component relevant to the pseudoscalar $B$
meson, $n_\perp=L=S=0$, and the transverse mode reduces to
$\phi_{0,0}(\zeta)= \kappa\sqrt{2} \zeta^{1/2}
e^{-\kappa^2\zeta^2/2}$.

In the massless limit, matching the electromagnetic and
gravitational form factors calculated in AdS space to their
light-front expressions determines the longitudinal mode as
$X(x)=\sqrt{x(1-x)}$ \cite{Brodsky2008a}. Substituting this
longitudinal profile together with the transverse ground-state mode
$\phi_{0,0}(\zeta)$ into Eq. (\ref{eq202}) yields the ground-state
holographic LFWF in transverse coordinate space,
\begin{equation}
\Psi(x,\mathbf{b}_\perp) =\frac{\kappa}{\sqrt{\pi}} \sqrt{x(1-x)}
\exp\left[ -\frac{\kappa^2x(1-x)\mathbf{b}_\perp^2}{2} \right],
\label{eq207}
\end{equation}
and, after Fourier transformation, in transverse momentum space,
\begin{equation}
\Psi(x,\mathbf{k}_\perp) =\frac{1}{\kappa\sqrt{\pi}}
\frac{1}{\sqrt{x(1-x)}} \exp\left[ -\frac{\mathbf{k}_\perp^2}
{2\kappa^2x(1-x)} \right]. \label{eq208}
\end{equation}
These expressions provide the transverse holographic structure that
will be retained in the heavy--light constructions considered below.

For a heavy--light meson such as the $B$-meson, the mass asymmetry
between the heavy quark and the light spectator generates a
nontrivial longitudinal structure that cannot be described
adequately by the massless form $X(x)=\sqrt{x(1-x)}$. We therefore
consider two complementary implementations of finite-quark-mass
effects. In the IMA, the massless light-front kinetic term appearing
in the momentum-space wave function is replaced by the full
invariant mass of the valence state \cite{Brodsky2014,Ahmady2018},
\begin{equation}
\frac{\mathbf{k}_\perp^2}{x(1-x)} \longrightarrow \mathcal{M}^2
=\frac{\mathbf{k}_\perp^2+m_q^2}{x} +
\frac{\mathbf{k}_\perp^2+m_b^2}{1-x}. \label{eq209}
\end{equation}
This prescription preserves the Gaussian transverse structure while
introducing the heavy--light asymmetry through the invariant-mass
dependence of the wave function.

As an alternative, the longitudinal mode can be determined
dynamically from an effective longitudinal light-front equation with
explicit quark-mass dependence. Such a construction is realized in
basis light-front quantization and in related longitudinally
improved holographic approaches \cite{Li2016,Jia2019}. In this case,
the longitudinal profile is obtained from the ground-state solution
of a confining longitudinal interaction rather than being introduced
solely through the invariant-mass substitution. The two approaches
therefore share the same transverse holographic structure used here
but encode the heavy--light longitudinal dynamics in different ways.
This distinction is central to the comparison developed in the
following section, where we construct the corresponding $B$-meson
LFWFs and derive their leading-twist LCDAs.

\section{Holographic $B$-Meson Light-Cone Distribution Amplitude}
\label{sec3}

We now construct the leading-twist, two-particle $B$-meson LCDA from
the holographic LFWFs. The heavy--light structure of the $B$-meson
makes this construction nontrivial because of the large mass
hierarchy $m_b \gg m_q$. Unlike light or equal-mass systems, for
which the longitudinal distribution is approximately symmetric, the
light spectator quark in a heavy meson is expected to carry only a
small fraction of the meson longitudinal momentum, whereas the heavy
quark carries the remaining fraction. A realistic holographic
description must therefore reproduce this strong longitudinal
asymmetry.

The leading two-particle LCDA in HQET, denoted by
$\phi_B^+(\omega)$, is a central nonperturbative input in
factorization analyses of exclusive heavy-meson decays. In the
present construction, we relate the dimensionless light-front
momentum fraction $x$ to the conventional HQET momentum variable
through the kinematic mapping $\omega = x m_B$.

The first inverse moment is defined as
\begin{equation}
\lambda_B^{-1} = \int_0^\infty d\omega\,
\frac{\phi_B^+(\omega)}{\omega}. \label{eq301}
\end{equation}
Because of the $1/\omega$ weighting, $\lambda_B$ is particularly
sensitive to the behavior of the LCDA near the soft endpoint,
$\omega\to0$, which corresponds to $x\to0$. A reliable description
of the longitudinal wave function in this region is therefore
essential for the determination of $\lambda_B$ and its
phenomenological applications.

In the following, we determine the leading-twist LCDA
$\phi_B^+(\omega)$ and its first inverse moment $\lambda_B$ within
the two holographic frameworks considered in this work. For both
models, the leading-twist DA is constructed directly from the
valence LFWF via transverse-momentum integration, followed by the
kinematic mapping $\omega = x m_B$ to the spectator momentum
variable. Since both approaches share the same transverse soft-wall
confinement scale $\kappa$, their comparison allows the effects of
the underlying longitudinal dynamics in the heavy-quark limit to be
examined while keeping the transverse confinement scale fixed. We
first analyze the kinematic prescription of the IMA, and
subsequently investigate the dynamical longitudinal confining
potential of the LMZV construction.

\subsection{Invariant mass ansatz}
\label{subseca}

First, the IMA approach is considered, which provides a direct
implementation of finite quark-mass effects in the holographic LFWF.
In this construction, the transverse dynamics remains governed by
the soft-wall confinement scale $\kappa$, while the heavy-light
asymmetry is introduced through the invariant mass of the valence
quark-antiquark state. This provides a particularly simple
realization of the longitudinal momentum hierarchy expected for a
$B$-meson, without introducing an additional longitudinal
interaction.

Starting from the massless momentum-space LFWF in Eq. (\ref{eq208})
and applying the invariant-mass replacement of Eq. (\ref{eq209}),
one obtains
\begin{equation}
\Psi_{B,\mathrm{IMA}}(x,\mathbf{k}_\perp) \propto
\frac{1}{\sqrt{x(1-x)}} \exp\left[ -\frac{\mathbf{k}_\perp^2}
{2\kappa^2x(1-x)} -\frac{m_q^2}{2\kappa^2x}
-\frac{m_b^2}{2\kappa^2(1-x)} \right]. \label{eq301a}
\end{equation}
The first term in the exponent retains the transverse Gaussian
structure of the soft-wall wave function, whereas the remaining
terms arise from the finite quark masses. In particular, the terms
$m_q^2/x$ and $m_b^2/(1-x)$ suppress the endpoint regions $x\to0$
and $x\to1$, respectively. Since $m_b\gg m_q$, the resulting
longitudinal distribution is strongly asymmetric and is concentrated
toward the small-$x$ region, as expected when $x$ denotes the
momentum fraction of the light spectator quark.

At the low hadronic scale $\mu_0$, the leading-twist DA associated
with the leading valence Fock sector is modeled through the
transverse-momentum integration of the valence LFWF
\cite{Brodsky1989,Lepage1980}:
\begin{equation}
\phi_{B}^+(x,\mu_0) = \int\frac{d^2\mathbf{k}_\perp}{16\pi^3}
\Psi_{B}(x,\mathbf{k}_\perp). \label{eq302a}
\end{equation}

Since the transverse dependence in Eq. (\ref{eq301a}) is Gaussian,
the integral can be evaluated analytically, yielding
\begin{equation}
\phi_{B,\mathrm{IMA}}^+(x) \propto \sqrt{x(1-x)} \exp\left[
-\frac{m_q^2}{2\kappa^2x} -\frac{m_b^2}{2\kappa^2(1-x)} \right].
\label{eq303a}
\end{equation}

As noted above, the relation $\omega=xm_B$ serves as a
phenomenological mapping between the partonic momentum fraction and
the HQET spectator variable, rather than an exact derivation of the
HQET light-cone momentum from the full-QCD light-front fraction.
Preservation of the normalization of the model LCDA under this
change of variables gives $\phi_{B,\mathrm{IMA}}^+(\omega)\,d\omega
= \phi_{B,\mathrm{IMA}}^+(x)\,dx$, and hence
\begin{equation}
\phi_{B,\mathrm{IMA}}^+(\omega) = \frac{1}{m_B}
\phi_{B,\mathrm{IMA}}^+ \left(x=\frac{\omega}{m_B}\right).
\label{eq304a}
\end{equation}

The factor $1/m_B$ is the Jacobian associated with the
transformation from the dimensionless partonic momentum fraction to
the dimensionful spectator-momentum variable. The original support
$0\leq x\leq1$ therefore corresponds to $0\leq\omega\leq m_B$ before
taking the heavy-quark approximation. Substituting $x=\omega/m_B$
into Eq. (\ref{eq303a}) and including the Jacobian in Eq.
(\ref{eq304a}) gives the finite-$m_B$ representation
\begin{equation}
\phi_{B,\mathrm{IMA}}^{+,\mathrm{finite}}(\omega) =
\frac{N_{\mathrm{IMA}}}{m_B} \left(\frac{\omega}{m_B}\right)^{1/2}
\left(1-\frac{\omega}{m_B}\right)^{1/2} \exp\left[ -\frac{A}{\omega}
-\frac{m_b^2} {2\kappa^2\left(1-\omega/m_B\right)} \right],
\label{eq305a}
\end{equation}
where $A=(m_q^2 m_B)/(2 \kappa^2)$, and $N_{\mathrm{IMA}}$ is a
normalization constant.

Keeping this expression explicit provides a convenient starting
point for assessing the accuracy of the subsequent heavy-quark
expansion. By expanding Eq. (\ref{eq305a}) in the region relevant
for the soft spectator, $\omega/m_B\ll1$, the prefactor becomes
\begin{equation}
\left(1-\frac{\omega}{m_B}\right)^{1/2} = 1-\frac{\omega}{2m_B}
+\mathcal{O}\left(\frac{\omega^2}{m_B^2}\right), \label{eq306a}
\end{equation}
while the heavy-quark term in the exponent satisfies
\begin{equation}
\left(1-\frac{\omega}{m_B}\right)^{-1} = 1+\frac{\omega}{m_B}
+\mathcal{O}\left(\frac{\omega^2}{m_B^2}\right). \label{eq307a}
\end{equation}
The constant term in Eq. (\ref{eq307a}) is independent of $\omega$
and can be absorbed into an overall normalization constant. The LP
approximation is obtained by retaining only the leading
$\omega$-dependent term in the exponent, which yields
\begin{equation}
\phi_{B,\mathrm{IMA}}^{+,\mathrm{LP}}(\omega) \propto \frac{1}{m_B}
\left(\frac{\omega}{m_B}\right)^{1/2} \exp\left[ -\frac{A}{\omega}
-B\omega \right], \label{eq308a}
\end{equation}
where $B=m_b^2/(2\kappa^2 m_B)$. At this stage, the physical support
$0\leq\omega\leq m_B$ is formally extended to $0\leq\omega<\infty$.
The LP LCDA model is normalized at the low hadronic scale $\mu_0$
according to
\begin{equation}
\int_0^\infty d\omega\,
\phi_{B,\mathrm{IMA}}^{+,\mathrm{LP}}(\omega) = 1. \label{eq309a}
\end{equation}
The required integral belongs to the standard class
\begin{equation}
\int_0^\infty d\omega\, \omega^{\nu-1} \exp\left(
-\frac{a}{\omega}-b\omega \right) =
2\left(\frac{a}{b}\right)^{\nu/2} K_\nu\left(2\sqrt{ab}\right),
\label{eq310a}
\end{equation}
where $K_\nu$ denotes the modified Bessel function of the second
kind. For the present LP model, the power of $\omega$ is
$\omega^{1/2}$, corresponding to $\nu=3/2$. Therefore,
\begin{equation}
\int_0^\infty d\omega\, \omega^{1/2} \exp\left(
-\frac{A}{\omega}-B\omega \right) = 2\left(\frac{A}{B}\right)^{3/4}
K_{3/2}\left(2\sqrt{AB}\right). \label{eq311a}
\end{equation}
The normalized LP LCDA is consequently
\begin{equation}
\phi_{B,\mathrm{IMA}}^{+,\mathrm{LP}}(\omega) = \frac{\omega^{1/2}
}{ 2 \left(\frac{A}{B}\right)^{3/4} K_{3/2}\left(2\sqrt{AB}\right) }
\exp\left[ -\frac{A}{\omega}-B\omega \right]. \label{eq312a}
\end{equation}

Thus, within the IMA construction at LP, the normalized LCDA is
determined by the effective quark masses, the transverse confinement
scale $\kappa$, and the meson mass $m_B$.

A careful distinction is required when assessing the heavy-quark
interpretation of the model. The IMA naturally produces the
soft-spectator region through the heavy-light invariant-mass
structure, and the resulting distribution is localized at
$\omega=\mathcal{O}(\Lambda_{\mathrm{QCD}})$. This endpoint
localization, however, does not by itself establish the complete
HQET scaling of all observables. In particular, the heavy-mass
dependence of the holographic parameters and of the light-front
current kernel must be controlled separately when testing the
asymptotic scaling of quantities such as the decay constant. We
therefore use Eq. (\ref{eq312a}) as the LP holographic LCDA model at
the physical $B$-meson point, while treating the finite-$m_B$
corrections and the full heavy-quark scaling of observables as
separate consistency checks.

Using Eq. (\ref{eq301}), the first inverse moment can be evaluated
with the same integral identity. Since $\phi_B^+(\omega)/\omega$ is
proportional to $\omega^{-1/2}\exp(-A/\omega-B\omega)$, the relevant
index is $\nu=1/2$. Hence,
\begin{equation}
\lambda_{B,\mathrm{IMA}}^{\mathrm{LP}} = \sqrt{\frac{A}{B}} \frac{
K_{3/2}\left(2\sqrt{AB}\right) }{ K_{1/2}\left(2\sqrt{AB}\right) }.
\label{eq313a}
\end{equation}

Thus, within the IMA model, $\lambda_{B,\mathrm{IMA}}^{\mathrm{LP}}$
is determined by the parameters entering the holographic wave
function rather than being introduced as an independent
phenomenological parameter. The factor $\exp(-A/\omega)$ determines
the suppression at small $\omega$, whereas $\exp(-B\omega)$ controls
the large-$\omega$ falloff of the leading-power model distribution.

\subsection{Longitudinal confinement in the LMZV framework}
\label{subsecb}

As a complementary approach to modeling the longitudinal dynamics of
the $B$-meson ($q\bar{b}$), the effective framework of LMZV
\cite{Li2016,Jia2019} is adopted. In the standard LFHQCD setup, the
holographic potential accounts exclusively for transverse
confinement through the dilaton term in Eq. (\ref{eq204}), leaving
the longitudinal momentum distribution $X(x)$ undetermined unless an
additional ansatz or heavy-quark constraint is imposed. To address
this limitation, the effective light-front Hamiltonian in Eq.
(\ref{eq201}) is approximated by assuming a separable interaction
kernel,
\begin{equation}
U(x,\mathbf{b}_\perp) \approx U_\perp(\zeta) + U_\parallel(x),
\label{eq301b}
\end{equation}
where $U_\perp(\zeta)$ retains the soft-wall harmonic form, while
$U_\parallel(x)$ governs the nonperturbative longitudinal
confinement.

While longitudinal confinement on the light front is originally
motivated by the $1+1$-dimensional 't Hooft model, in the LMZV
framework \cite{Li2016} it is modeled by an analytically solvable
differential operator acting on the longitudinal momentum fraction
$x$,
\begin{equation}
U_{\parallel}(x) = -\sigma^2\frac{d}{dx} \left[ x(1-x)\frac{d}{dx}
\right], \label{eq302b}
\end{equation}
where $\sigma$ denotes the characteristic longitudinal confinement
scale. Under the factorized representation of the LFWF in Eq.
(\ref{eq202}), and by factoring out the standard kinematical
phase-space prefactor, $X(x)=\sqrt{x(1-x)}\chi(x)$, the two-body
bound-state equation separates. The resulting longitudinal
eigenvalue equation for $\chi(x)$ takes the self-adjoint form
\begin{equation}
\left[ -\sigma^2\frac{d}{dx} \left( x(1-x)\frac{d}{dx} \right)
+\frac{m_q^2}{x} +\frac{m_b^2}{1-x} \right]\chi(x) = M_L^2\chi(x),
\label{eq303b}
\end{equation}
where $M_L^2$ is the longitudinal eigenvalue contribution to the
squared meson mass. In this equation, the first differential term
acts as an effective kinetic confinement barrier that suppresses
rapid variations in the longitudinal wave function, while the
mass-dependent terms $m_q^2/x$ and $m_b^2/(1-x)$ impose kinematic
endpoint penalties, leading to suppressed wave functions near
$x\to0$ and $x\to1$.

The eigenfunctions of the longitudinal operator can be expressed in
terms of Jacobi polynomials. For the present heavy--light system,
they take the form
\begin{equation}
\chi_{n_\parallel}(x) = N_{n_\parallel} x^{\alpha_0/2}
(1-x)^{\beta_0/2} P_{n_\parallel}^{(\beta_0,\alpha_0)}(2x-1),
\label{eq304b}
\end{equation}
where $\alpha_0=2m_q/\sigma$, and $\beta_0=2m_b/\sigma$. The two
indices determine the endpoint behavior of the longitudinal mode. In
particular, the heavy--light hierarchy $m_b\gg m_q$ implies
$\beta_0\gg\alpha_0$, leading to a strongly asymmetric longitudinal
distribution with enhanced support in the small-$x$ region relevant
for the light spectator.

For the physical $B$-meson, the longitudinal ground state
corresponding to $n_\parallel=0$ is retained. Since
$P_0^{(\beta_0,\alpha_0)}(2x-1)=1$, combining the lowest
longitudinal mode with the soft-wall transverse ground state in Eq.
(\ref{eq207}) yields the coordinate-space LFWF
\begin{equation}
\Psi_{B,\mathrm{LMZV}}(x,\mathbf{b}_\perp) \propto
x^{(\alpha_0+1)/2} (1-x)^{(\beta_0+1)/2} \exp\left[
-\frac{\kappa^2x(1-x) \mathbf{b}_\perp^2}{2} \right]. \label{eq305b}
\end{equation}
The transverse dependence thus retains the Gaussian form generated
by the soft-wall dynamics, while the heavy--light asymmetry resides
in the longitudinal exponents. Taking the two-dimensional Fourier
transform with respect to $\mathbf{b}_\perp$, one arrives at the
momentum-space wave function
\begin{equation}
\Psi_{B,\mathrm{LMZV}}(x,\mathbf{k}_\perp)\propto x^{(\alpha_0-1)/2}
(1-x)^{(\beta_0-1)/2} \exp\left[ -\frac{\mathbf{k}_\perp^2}
{2\kappa^2x(1-x)} \right]. \label{eq306b}
\end{equation}

The leading-twist LCDA is obtained from the valence LFWF by
integrating over transverse momentum, as given in Eq.
(\ref{eq302a}). Since
\begin{equation}
\int d^2\mathbf{k}_\perp \exp\left[ -\frac{\mathbf{k}_\perp^2}
{2\kappa^2x(1-x)} \right] = 2\pi\kappa^2x(1-x), \label{eq307b}
\end{equation}
the transverse integration leaves the resulting LCDA proportional to
the ground-state mode in Eq. (\ref{eq304b}) together with the
corresponding phase-space factor. Following the same normalization
procedure as in the IMA construction, the normalized leading-twist
LCDA is given by
\begin{equation}
\phi_{B,\mathrm{LMZV}}^+(x) = \frac{ x^{(\alpha_0+1)/2}
(1-x)^{(\beta_0+1)/2} }{ B\left( \frac{\alpha_0+3}{2},
\frac{\beta_0+3}{2} \right) }, \label{eq308b}
\end{equation}
where $B(a,b)$ denotes the Euler beta function. Unlike the IMA
construction, where the longitudinal profile is generated through
the invariant-mass dependence of the holographic wave function, the
endpoint exponents in Eq. (\ref{eq308b}) arise directly from the
dynamical longitudinal confining interaction and are governed by
$\alpha_0$ and $\beta_0$.

The phenomenological mapping $\omega=xm_B$, which connects the model
momentum fraction to the HQET spectator variable, yields the
finite-$m_B$ representation:
\begin{equation}
\phi_{B,\mathrm{LMZV}}^{+,\mathrm{finite}}(\omega) = \frac{
\left(\frac{\omega}{m_B}\right)^{(\alpha_0+1)/2}
\left(1-\frac{\omega}{m_B}\right)^{(\beta_0+1)/2} }{ m_B B\left(
\frac{\alpha_0+3}{2}, \frac{\beta_0+3}{2} \right) }. \label{eq309b}
\end{equation}

Retaining this finite-$m_B$ expression provides a transparent
starting point for assessing the accuracy of the subsequent
heavy-quark expansion. In the kinematic domain relevant for the soft
spectator, $\omega/m_B \ll 1$, the heavy-quark endpoint factor can
be expanded systematically as
\begin{align}
\left(1-\frac{\omega}{m_B}\right)^{(\beta_0+1)/2} &= \exp\left[
\frac{\beta_0+1}{2} \ln\left(1-\frac{\omega}{m_B}\right) \right] \nonumber \\
&= \exp\left[ -\left(\frac{m_b}{\sigma}+\frac{1}{2}\right) \left(
\frac{\omega}{m_B} + \frac{\omega^2}{2m_B^2} +
\mathcal{O}\left(\frac{\omega^3}{m_B^3}\right) \right) \right] \nonumber \\
&= \exp\left(-\frac{\omega}{\sigma}\right) \left[ 1-
\frac{\omega}{2m_B} \left( 1+\frac{\omega}{\sigma} \right) +
\mathcal{O}\left(\frac{\omega^2}{m_B^2}\right) \right],
\label{eq310b}
\end{align}
where the heavy-quark relation $m_B \approx m_b$ has been used.
Concurrently, the asymptotic behavior of the normalization factor is
governed by the large-$b$ expansion of the Beta function. Defining
$a=(\alpha_0+3)/2$, and $b=(\beta_0+3)/2 = m_b/\sigma+3/2$, and
using
\begin{equation}
B(a,b) = \frac{\Gamma(a)\Gamma(b)} {\Gamma(a+b)} = \Gamma(a)b^{-a}
\left[ 1+\mathcal{O}\left(\frac{1}{b}\right) \right], \label{eq311b}
\end{equation}
one obtains
\begin{equation}
m_B B\left( \frac{\alpha_0+3}{2}, \frac{\beta_0+3}{2} \right) =
\Gamma\left( \frac{\alpha_0+3}{2} \right) \sigma^{(\alpha_0+3)/2}
m_b^{1-(\alpha_0+3)/2} \left[
1+\mathcal{O}\left(\frac{\sigma}{m_b}\right) \right], \label{eq312b}
\end{equation}
where $m_B\simeq m_b$ has been used at leading order. The explicit
$m_b$ dependence in Eq. (\ref{eq312b}) cancels against the
corresponding factor from $(\omega/m_B)^{(\alpha_0+1)/2}$ in the
numerator of Eq. (\ref{eq309b}). The leading corrections arising
from both the endpoint factor and the normalization are therefore of
relative order $\Lambda_{\mathrm{QCD}}/m_b$ in the soft-spectator
region $\omega\sim\Lambda_{\mathrm{QCD}}$. Retaining only the LP
terms in the heavy-quark expansion and extending the support domain
from $0\leq\omega\leq m_B$ to the semi-infinite interval
$0\leq\omega<\infty$, we arrive at the LP LCDA:
\begin{equation}
\phi_{B,\mathrm{LMZV}}^{+,\mathrm{LP}}(\omega) = \frac{
\omega^{(\alpha_0+1)/2} e^{-\omega/\sigma} }{
\sigma^{(\alpha_0+3)/2} \Gamma\left( \frac{\alpha_0+3}{2} \right) }.
\label{eq313b}
\end{equation}
This LP distribution satisfies the standard normalization condition
$\int_0^\infty d\omega\,
\phi_{B,\mathrm{LMZV}}^{+,\mathrm{LP}}(\omega) = 1$, and has a
Gamma-type functional form with an exponentially suppressed
large-$\omega$ tail set by the longitudinal confinement scale
$\sigma$.

Evaluating the inverse moment defined in Eq. (\ref{eq301}) for the
finite-$m_B$ distribution in Eq. (\ref{eq309b}) yields
\begin{equation}
\left(\lambda_{B,\mathrm{LMZV}}^{\mathrm{finite}}\right)^{-1} =
\frac{1}{m_B} \frac{ B\left( \frac{\alpha_0+1}{2},
\frac{\beta_0+3}{2} \right) }{ B\left( \frac{\alpha_0+3}{2},
\frac{\beta_0+3}{2} \right) }. \label{eq314b}
\end{equation}
Using the recurrence relation for the Beta function,
$B(a,b)/B(a+1,b)=(a+b)/a$, this reduces to
\begin{equation}
\lambda_{B,\mathrm{LMZV}}^{\mathrm{finite}} = m_B \frac{\alpha_0+1}
{\alpha_0+\beta_0+4}. \label{eq315b}
\end{equation}

Within this construction, $\lambda_B$ is determined by the
parameters of the longitudinal confining potential and the
constituent quark masses, rather than being introduced as an
independent phenomenological fit parameter. In the heavy-quark limit
$m_b\to\infty$, substituting $\alpha_0=2m_q/\sigma$ and
$\beta_0=2m_b/\sigma$ along with $m_B\approx m_b$ gives
\begin{align}
\lambda_{B,\mathrm{LMZV}}^{\mathrm{finite}} = m_B
\frac{2m_q/\sigma+1} {2(m_q+m_b)/\sigma+4} = \left(
m_q+\frac{\sigma}{2} \right) \left[ 1+ \mathcal{O}\left(
\frac{\Lambda_{\mathrm{QCD}}}{m_b} \right) \right], \label{eq316b}
\end{align}
where the last expression represents the leading heavy-quark
behavior. Taking the strict heavy-quark limit yields the LP inverse
moment
\begin{equation}
\lambda_{B,\mathrm{LMZV}}^{\mathrm{LP}} = m_q+\frac{\sigma}{2},
\label{eq317b}
\end{equation}
which is reproduced by direct integration of the LP LCDA
$\phi_{B,\mathrm{LMZV}}^{+,\mathrm{LP}}(\omega)$ over the domain
$[0,\infty)$. Equation (\ref{eq317b}) makes transparent that the
inverse moment is of order $\mathcal{O}(\Lambda_{\mathrm{QCD}})$ in
the heavy-quark limit, with the leading heavy-quark dependence
canceling in this construction.

The LMZV and IMA constructions thus implement the heavy--light
longitudinal dynamics in qualitatively different ways. In the IMA
approach, the quark masses enter directly through the invariant mass
of the valence Fock state, which generates the soft-spectator
enhancement and leads, upon taking the LP heavy-quark limit, to the
Bessel-regulated distribution
$\phi_{B,\mathrm{IMA}}^{+,\mathrm{LP}}(\omega)$ in Eq.
(\ref{eq312a}) with an essential inverse-$\omega$ exponential
factor. In the LMZV construction, the longitudinal dependence is
generated dynamically by the effective longitudinal confining
operator, yielding the Beta-type distribution
$\phi_{B,\mathrm{LMZV}}^{+,\mathrm{finite}}(\omega)$ in Eq.
(\ref{eq309b}) on the kinematically bounded domain $0\leq\omega\leq
m_B$. In the heavy-quark limit $m_b\to\infty$, the latter
asymptotically reduces to the Gamma-type distribution
$\phi_{B,\mathrm{LMZV}}^{+,\mathrm{LP}}(\omega)$ given in Eq.
(\ref{eq313b}).

The two constructions therefore provide complementary holographic
realizations of the same heavy--light system while differing in
their microscopic origin of longitudinal confinement. Comparing
these frameworks allows us to examine how the underlying
longitudinal dynamics affects both the functional shape of the
$B$-meson LCDA and the phenomenologically relevant first inverse
moment $\lambda_B$. In both approaches, $\lambda_B$ is determined by
the parameters of the corresponding holographic wave function [Eq.
(\ref{eq313a}) for IMA and Eq. (\ref{eq317b}) for LMZV], rather than
introduced as an independent fit parameter.

\subsection{Inputs and the $B$-meson LCDA}
\label{secc}

The construction of the $B$-meson LCDA within the light-front
holographic framework requires a consistent specification of the
nonperturbative inputs governing the transverse and longitudinal
dynamics. In the present analysis, these inputs are the transverse
confinement scale $\kappa$, the longitudinal confinement scale
$\sigma$, and the effective mass $m_q$ of the light spectator quark.
These parameters enter different components of the effective
light-front Hamiltonian and therefore play distinct physical roles.

For the longitudinal dynamics, the LMZV interaction is employed,
which provides a dynamical description of longitudinal confinement
and incorporates finite-quark-mass effects beyond the IMA. The
eigenfunctions of the longitudinal LMZV equation, their endpoint
behavior, and their characteristic momentum scale depend on the
dimensionless ratios $m_q/\sigma$ and $m_b/\sigma$. Consequently,
the choice of $\sigma$ affects not only the longitudinal
contribution to the meson mass but also the shape of the
longitudinal wave function and, through it, the resulting $B$-meson
LCDA \cite{Li2022,DeTeramond2021}.

The effective transverse confinement scale in light-front holography
is not universal across different sectors. For light hadrons, a
reference value of $\kappa \simeq 0.523 \mathrm{GeV}$ is consistent
with the Regge slopes \cite{Brodsky2016,DeTeramond2014}, whereas
this parameter varies in the presence of heavy flavors
\cite{Chang2017,Chang2017a}. For the pseudoscalar $B$-meson, an
effective transverse confinement scale of $\kappa = 0.590 \pm 0.050
\mathrm{GeV}$ is adopted for the present phenomenological analysis.
The central value is moderately larger than the light-hadron
reference scale, while the quoted interval is treated as a model
variation and is propagated through the numerical analysis.

The longitudinal confinement scale $\sigma$ requires dedicated
treatment, as its value depends on the implementation of the
longitudinal potential and the relevant quark-mass regime. In the
heavy-quark limit of LFHQCD, rotational invariance yields the
matching relation $\sigma \simeq \lambda_Q / M_M$
\cite{Nielsen2018,DeTeramond2021}. Combined with the heavy-quark
scaling $\sqrt{\lambda_Q} = C \sqrt{M_M}$ with $C = (0.49 \pm 0.02)
\mathrm{GeV}^{1/2}$, this gives the benchmark estimate $\sigma
\simeq C^2 \simeq 0.24 \mathrm{GeV}$. Conversely, alternative
holographic implementations assume a universal $\sigma = 0.12
\mathrm{GeV}$ across all meson sectors while varying the transverse
scale $\kappa$ \cite{Ahmady2021}.

Given that the heavy-light $B$-meson is intrinsically asymmetric,
its effective longitudinal dynamics need not coincide with either
extreme regime. Accordingly, $\sigma$ is treated here as an
effective heavy-light input rather than a universally constrained
scale. We adopt the phenomenological intermediate value $\sigma =
0.200 \pm 0.040 \mathrm{GeV}$, which spans the interval $0.16 \le
\sigma \le 0.24 \mathrm{GeV}$ between these literature benchmarks
\cite{Ahmady2021,DeTeramond2021}. This range reflects a
model-parameter variation to assess the sensitivity of observables
to the longitudinal dynamics, rather than a statistical uncertainty
from a global fit.

In light-front holographic approaches, the light-quark masses
entering the effective LFWF are understood as effective parameters
associated with the reduction of higher Fock-state degrees of
freedom and therefore need not coincide with the current quark
masses appearing in the QCD Lagrangian \cite{Brodsky2014}. Moreover,
phenomenological light-front calculations of heavy-light mesons
employ light-quark masses of constituent-like magnitude. In the
present numerical analysis, we therefore take $m_q = 0.300 \pm 0.030
\mathrm{GeV}$ \cite{Cheung1996} as an effective mass for the light
spectator quark in the heavy-light bound state. Since the
longitudinal wave function depends on the combination $m_q/\sigma$
through $\alpha_0 = 2m_q/\sigma$, this choice directly influences
the longitudinal endpoint behavior and the resulting momentum
distribution. As in the cases of $\kappa$ and $\sigma$, the quoted
interval is treated as a phenomenological model variation rather
than as a statistical uncertainty.

\begin{figure*}[t]
\centering \subfigure[]{
\includegraphics[width=0.48\textwidth]{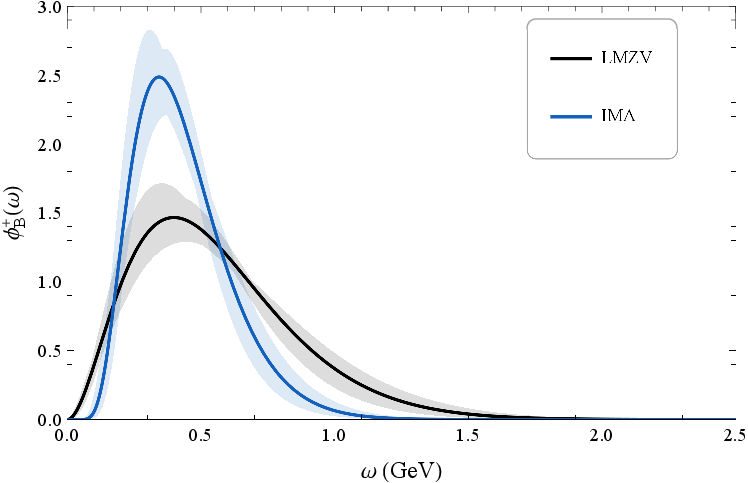}
} \hfill \subfigure[]{
\includegraphics[width=0.48\textwidth]{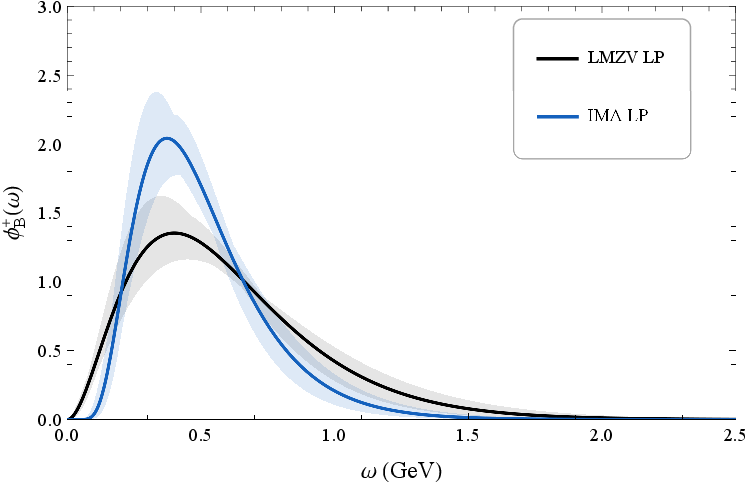}
}\caption{The leading-twist $B$-meson LCDA $\phi_B^+(\omega)$
obtained in the LMZV (solid black curves) and IMA (solid blue
curves) frameworks. Specifically, (a) shows the finite-$m_B$
distributions given by Eqs. (\ref{eq309b}) and (\ref{eq305a}), while
(b) displays the corresponding LP asymptotic limits from Eqs.
(\ref{eq313b}) and (\ref{eq312a}). The shaded bands indicate the
parametric variations induced by $m_q$ and $\kappa$ in the IMA
framework and by $m_q$ and $\sigma$ in the LMZV framework.}
\label{F31}
\end{figure*}

The resulting leading-twist $B$-meson LCDAs are displayed in Fig.
\ref{F31}, where (a) illustrates the finite-$m_B$ distributions and
(b) shows the corresponding LP asymptotic distributions. For the
central input parameters, the finite-$m_B$ distributions peak at
approximately $\omega\simeq0.34 \mathrm{GeV}$ in the IMA framework
and $\omega\simeq0.40 \mathrm{GeV}$ in the LMZV framework. The
corresponding LP distributions peak at approximately $0.37
\mathrm{GeV}$ and $0.40 \mathrm{GeV}$, respectively. Variations of
the model parameters shift these peak positions. However, the IMA
distribution exhibits a noticeably narrower and steeper peak
compared to the broader, dynamically confined LMZV counterpart. This
feature reflects the different endpoint suppression mechanisms: the
inverse-power exponential factor in the IMA concentrates the
momentum fraction more strongly than the power-law and exponential
behavior governed by the LMZV Jacobi parameters.

Furthermore, as summarized in Table \ref{T31}, the first inverse
moment $\lambda_B$ obtained in the LMZV framework shifts from
$\lambda_{B,\mathrm{LMZV}}^{\mathrm{finite}} =
0.384^{+0.040}_{-0.041} \mathrm{GeV}$ to
$\lambda_{B,\mathrm{LMZV}}^{\mathrm{LP}} = 0.400^{+0.050}_{-0.050}
\mathrm{GeV}$ in the heavy-quark limit, representing a modest $\sim
4\%$ power correction. In contrast, the IMA inverse moment exhibits
a larger relative shift of $\sim 13\%$, indicating a greater
sensitivity of the IMA result to the heavy-quark expansion.
\begin{table}[t]
\centering \caption{Comparison of the first inverse moment
$\lambda_B$ in the LMZV and IMA frameworks, for both finite-$m_B$
distributions and their LP limits, with parametric uncertainties.}
\label{T31}
\begin{ruledtabular}
\begin{tabular}{lc}
Model & $\lambda_B\text{ [GeV]}$ \\
\hline
LMZV (finite-$m_B$) & $0.384^{+0.040}_{-0.041}$ \\[0.5ex]
LMZV LP             & $0.400^{+0.050}_{-0.050}$ \\[0.5ex]
IMA (finite-$m_B$)  & $0.364^{+0.036}_{-0.036}$ \\[0.5ex]
IMA LP              & $0.410^{+0.047}_{-0.046}$ \\
\end{tabular}
\end{ruledtabular}
\end{table}

\section{Heavy-Quark Scaling and Consistency with HQET}
\label{sec4}

A fundamental consistency test for any relativistic light-front
model of heavy--light mesons is its compliance with the asymptotic
scaling laws predicted by heavy-quark effective theory (HQET)
\cite{Isgur1989,Neubert1994}. In the static heavy-quark limit ($m_b
\to \infty$), the light-quark dynamics decouples from the
heavy-quark spin and flavor degrees of freedom. In particular, the
pseudoscalar meson decay constant satisfies the universal power-law
scaling \cite{Neubert1994}
\begin{equation}
f_B \sqrt{m_B} = C_{\mathrm{HQET}}\left[1 +
\mathcal{O}\!\left(\frac{\Lambda_{\mathrm{QCD}}}{m_B}\right)\right],
\qquad m_B \to \infty, \label{eq401}
\end{equation}
up to perturbative matching logarithms.

In this regime, the support of the soft spectator momentum fraction
is concentrated in the region $x \sim \Lambda_{\mathrm{QCD}}/m_B \ll
1$, while the hadronic momentum variable $\omega = x m_B$ remains of
order $\mathcal{O}(\Lambda_{\mathrm{QCD}})$. When examining the
asymptotic consistency of holographic LFWFs, it is essential to
distinguish two independent criteria:
\begin{enumerate}
    \item The longitudinal mode should retain a non-singular
    soft profile with a finite width of order
    $\mathcal{O}(\Lambda_{\mathrm{QCD}})$ in $\omega$ as $m_b\to\infty$.
    Although a finite inverse moment
    $\lambda_B=\mathcal{O}(\Lambda_{\mathrm{QCD}})$ is a relevant
    requirement, it does not by itself guarantee that the
    longitudinal distribution retains a nontrivial hadronic width.
    \item The local-current matrix element must reproduce the
    $f_B\sim m_B^{-1/2}$ behavior of
    Eq. (\ref{eq401}) for a physical prediction,
    with the appropriate short-distance current matching taken into
    account. The scaling obtained from the bare light-front current
    is therefore examined separately as a diagnostic of the
    current-matrix-element formalism.
\end{enumerate}

In what follows, we systematically assess both criteria for the IMA
and LMZV frameworks.

\subsection{Longitudinal Dynamics and Heavy-Quark Scaling Consistency}
\label{subsecaa}

Within light-front bound-state frameworks, the behavior of
heavy--light mesons in the static limit $m_b \to \infty$ depends
sensitively on the longitudinal component of the valence wave
function. Before integrating over the transverse momentum to extract
the LCDA $\phi_B^+(\omega)$, analyzing $X(x)$ directly offers useful
insight into the underlying heavy-quark scaling. If the transverse
dependence remains governed by a fixed soft confinement scale,
$\kappa \sim \Lambda_{\mathrm{QCD}}$, the resulting LCDA inherits
the dominant longitudinal localization properties. We therefore
examine the heavy-quark behavior of the longitudinal dynamics in the
IMA and LMZV frameworks.

To clarify the asymptotic behavior of the longitudinal mode within
the IMA, it is instructive to examine both its peak position and
effective hadronic width directly in terms of the HQET spectator
momentum $\omega = x m_B$. In the IMA, the asymmetry in longitudinal
momentum is governed kinematicly by the two-particle invariant mass,
\begin{equation}
M_0^2(x) = \frac{m_q^2}{x} + \frac{m_b^2}{1-x}, \label{eq401a}
\end{equation}
which enters the wave function through the Gaussian factor
$\exp[-M_0^2(x)/(2\kappa^2)]$. In the soft regime with $\omega =
\mathcal{O}(\Lambda_{\mathrm{QCD}})$ and $m_b \to \infty$, utilizing
the expansion $(1-\omega/m_B)^{-1} = 1 + \omega/m_B +
\mathcal{O}(\omega^2/m_B^2)$, the two-particle invariant mass in Eq
. (\ref{eq401a}) can be expanded as
\begin{equation}
M_0^2(\omega) = m_b^2 + \frac{m_B m_q^2}{\omega} +
\frac{m_b^2}{m_B}\omega + \mathcal{O}(\Lambda_{\mathrm{QCD}}^2).
\label{eq402a}
\end{equation}
The $\omega$-dependence of the Gaussian factor
$\exp[-M_0^2/(2\kappa^2)]$ can be expressed as
$\exp[-\Phi_{\mathrm{IMA}}(\omega)]$, where the effective exponent
is defined by
\begin{equation}
\Phi_{\mathrm{IMA}}(\omega) = \frac{1}{2\kappa^2} \left[ \frac{m_B
m_q^2}{\omega} + \frac{m_b^2}{m_B}\omega \right] + \mathcal{O}\left(
\frac{\Lambda_{\mathrm{QCD}}^2}{\kappa^2} \right). \label{eq403a}
\end{equation}
The stationary condition $\Phi_{\mathrm{IMA}}'(\omega) = 0$
determines the peak location of the dominant exponential:
\begin{equation}
\omega_0 = \frac{m_B}{m_b} m_q = m_q +
\mathcal{O}\left(\frac{\Lambda^2_{\mathrm{QCD}}}{m_b}\right).
\label{eq404a}
\end{equation}
Thus, the longitudinal mode remains centered at a soft momentum
scale, $\omega_0 = \mathcal{O}(\Lambda_{\mathrm{QCD}})$.

However, its characteristic hadronic width decreases in the
heavy-quark limit. Expanding $\Phi_{\mathrm{IMA}}(\omega)$ around
the minimum $\omega_0 \simeq m_q$, the curvature satisfies
$\Phi_{\mathrm{IMA}}''(m_q) \simeq m_B / (\kappa^2 m_q)$.
Consequently, the effective momentum width around the peak decreases
as
\begin{equation}
\Delta\omega \sim \left[ \Phi_{\mathrm{IMA}}''(m_q) \right]^{-1/2}
\sim \kappa \sqrt{\frac{m_q}{m_B}} \propto m_B^{-1/2}
\xrightarrow{m_B \to \infty} 0. \label{eq405a}
\end{equation}

More precisely, upon transforming to the momentum variable $\omega$
and imposing unit normalization, the longitudinal distribution
becomes progressively narrower, and its LP counterpart
$\phi_{B,\mathrm{IMA}}^{+,\mathrm{LP}}(\omega)$ approaches a
singular limiting form concentrated at the soft scale $\omega_0 \to
m_q$:
\begin{equation}
\phi_{B,\mathrm{IMA}}^{+,\mathrm{LP}}(\omega)
\xrightarrow{m_B\to\infty} \delta(\omega-m_q), \label{eq406a}
\end{equation}
in the sense of generalized functions. The corresponding inverse
moment in Eq. (\ref{eq313a}) therefore satisfies
\begin{equation}
\left(\lambda_{B,\mathrm{IMA}}^{\mathrm{LP}}\right)^{-1} =
\int_0^\infty \frac{d\omega}{\omega} \,
\phi_{B,\mathrm{IMA}}^{+,\mathrm{LP}}(\omega)
\xrightarrow{m_B\to\infty} \frac{1}{m_q}, \qquad
\lambda_{B,\mathrm{IMA}}^{\mathrm{LP}} \xrightarrow{m_B\to\infty}
m_q = \mathcal{O}(\Lambda_{\mathrm{QCD}}). \label{eq407a}
\end{equation}
Thus, the characteristic limitation of the IMA in the heavy-quark
limit is not a vanishing inverse moment. Rather, at fixed soft
confinement scale, the normalized distribution loses its finite
hadronic width and becomes parametrically narrow in $\omega$.
Maintaining a finite width within this scaling estimate would
require $\kappa^2\propto m_B$, or $\kappa\propto\sqrt{m_B}$, rather
than a fixed soft confinement scale
$\kappa\sim\Lambda_{\mathrm{QCD}}$.

In contrast, the LMZV framework generates the longitudinal wave
function dynamically through the confining potential
$U_\parallel(x)$. In terms of the Jacobi ground-state mode in Eq.
(\ref{eq304b}) modulated by the effective measure factor
$\sqrt{x(1-x)}$, the longitudinal wave function reads
\begin{equation}
X_0(x) = N_0\, x^{(\alpha_0+1)/2} (1-x)^{(\beta_0+1)/2},
\label{eq408a}
\end{equation}
where the invariant phase-space normalization $\int_0^1
\frac{dx}{x(1-x)} |X_0(x)|^2 = 1$ determines the normalization
constant as $N_0 = [B(\alpha_0+1, \beta_0+1)]^{-1/2}$.

Using the mapping $\omega=xm_B$ to express the longitudinal mode in
terms of the spectator momentum gives
\begin{equation}
X_0\left(\frac{\omega}{m_B}\right) = \frac{1}{\sqrt{B(\alpha_0+1,
\beta_0+1)}} \left(\frac{\omega}{m_B}\right)^{(\alpha_0+1)/2}
\left(1 - \frac{\omega}{m_B}\right)^{(\beta_0+1)/2}. \label{eq409a}
\end{equation}

The peak position $\omega_0$ is obtained from the stationary
condition $\frac{d}{d\omega} X_0(\omega/m_B) = 0$. Using the LMZV
parameter relations $\alpha_0 = 2m_q/\sigma$ and $\beta_0 =
2m_b/\sigma$, this condition yields
\begin{equation}
\omega_0 = m_B \frac{\alpha_0+1}{\alpha_0+\beta_0+2} = m_B
\frac{m_q+\sigma/2}{m_b+m_q+\sigma} \xrightarrow{m_b\to\infty} m_q +
\frac{\sigma}{2} = \mathcal{O}(\Lambda_{\mathrm{QCD}}).
\label{eq410a}
\end{equation}
Thus, the peak remains at a soft hadronic scale in the heavy-quark
limit.

The large-$\omega$ falloff of Eq. (\ref{eq409a}) is governed by the
binomial factor $(1 - \omega/m_B)^{(\beta_0+1)/2}$. In the
heavy-quark limit with $\beta_0 \approx 2m_b/\sigma \to \infty$,
this factor approaches an exponential form:
\begin{equation}
\lim_{m_b\to\infty} \left(1 -
\frac{\omega}{m_B}\right)^{(\beta_0+1)/2} = e^{-\omega/\sigma}.
\label{eq411a}
\end{equation}
Because the falloff rate is controlled by the mass-independent
parameter $\sigma \sim \Lambda_{\mathrm{QCD}}$, the characteristic
hadronic width remains of order $\sigma$:
\begin{equation}
\lim_{m_B \to \infty} \Delta\omega \sim \sigma =
\mathcal{O}(\Lambda_{\mathrm{QCD}}) \neq 0. \label{eq412a}
\end{equation}

The limiting functional shape of the longitudinal mode is obtained
by evaluating the normalization constant in the heavy-quark limit,
\begin{equation}
N_0 \simeq
\frac{\left(\frac{2m_b}{\sigma}\right)^{(\alpha_0+1)/2}}{\sqrt{\Gamma(\alpha_0+1)}}
\propto m_b^{(\alpha_0+1)/2}. \label{eq413a}
\end{equation}
The kinematic suppression factor $m_B^{-(\alpha_0+1)/2}$ in Eq.
(\ref{eq409a}) is precisely canceled by the asymptotic growth of
$N_0$. Consequently, the longitudinal mode approaches a stable LP
distribution:
\begin{equation}
X_{0}\left(\frac{\omega}{m_B}\right) \xrightarrow{m_b \to \infty}
\mathcal{X}_{\mathrm{LP}}(\omega) =
\frac{(2\omega/\sigma)^{(\alpha_0+1)/2}
e^{-\omega/\sigma}}{\sqrt{\Gamma(\alpha_0+1)}} = \mathcal{O}(1),
\label{eq414a}
\end{equation}
without the singular delta-function limit found in the IMA.

Finally, in accordance with Eq. (\ref{eq317b}), the LP
inverse-moment parameter $\lambda_B$ satisfies
\begin{equation}
\lambda_{B,\mathrm{LMZV}}^{\mathrm{LP}} \xrightarrow{m_B\to\infty}
m_q + \frac{\sigma}{2} = \mathcal{O}(\Lambda_{\mathrm{QCD}}).
\label{eq415a}
\end{equation}
Thus, within the LMZV construction, the dynamical longitudinal
potential preserves a stable hadronic width $\Delta\omega \sim
\sigma$ and a non-singular soft profile in the heavy-quark limit,
without requiring a mass-dependent scaling of the confinement scale.

In summary, the main distinction between the two frameworks does not
lie in the parametric order of $\lambda_B$, which remains soft in
both constructions, but in the heavy-quark-limit behavior of the
distribution. For fixed soft $\kappa$, the IMA yields a narrowing
width $\Delta\omega \sim m_B^{-1/2} \to 0$, whereas the dynamical
longitudinal confinement of the LMZV framework retains a finite
hadronic width $\Delta\omega \sim \sigma \sim
\Lambda_{\mathrm{QCD}}$ in the heavy-quark limit.

\begin{figure*}[t]
\centering
\begin{minipage}[t]{0.48\textwidth}
\centering
\includegraphics[width=\linewidth]{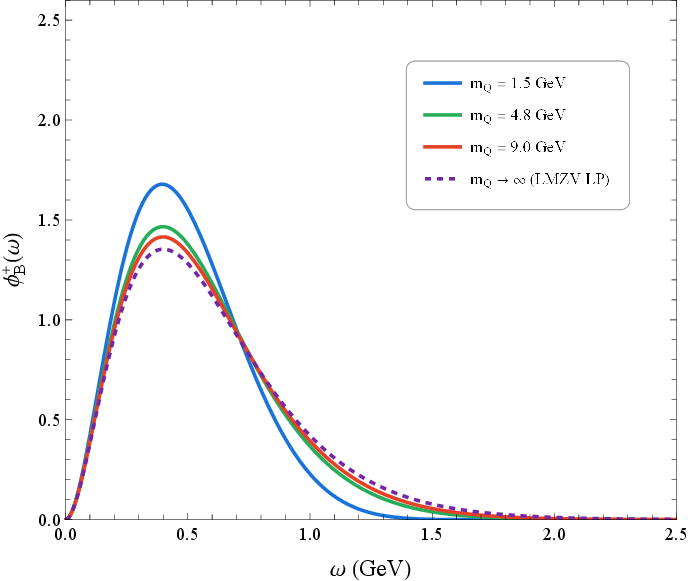}
\\ (a)
\end{minipage}\hfill
\begin{minipage}[t]{0.48\textwidth}
\centering
\includegraphics[width=\linewidth]{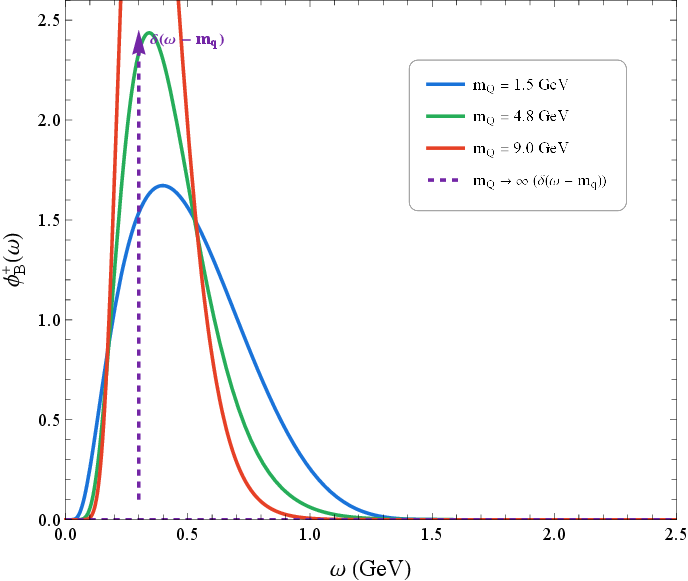}
\\ (b)
\end{minipage}
\caption{Evolution of the leading-twist LCDA $\phi_B^+(\omega)$ for
heavy--light mesons with increasing heavy-quark mass $m_Q \in \{1.5,
4.8, 9.0\} \mathrm{GeV}$ alongside the static limit $m_Q \to
\infty$: (a) the LMZV model, and (b) the IMA approach.} \label{F41}
\end{figure*}

The analytic behavior derived above for the longitudinal mode
$X_0(\omega)$ can be compared with the full LCDA $\phi_B^+(\omega)$
displayed in Fig. \ref{F41}, which is obtained upon integration over
the transverse momentum.
Consistent with the asymptotic functional form
$\mathcal{X}_{\mathrm{LP}}(\omega)$ in Eq. (\ref{eq414a}), the LMZV
LCDAs in (a) approach a stable LP form with a finite width of order
$\sigma$. In contrast, the IMA counterparts in (b) become
progressively taller and narrower as $m_Q$ increases, reflecting the
width scaling $\Delta\omega \propto m_Q^{-1/2}$ of Eq.
(\ref{eq405a}) and their distributional limit $\delta(\omega -
m_q)$. This behavior is consistent with the asymptotic analysis of
the longitudinal dynamics and is retained after transverse
integration within the model construction.

\subsection{Current Matrix Elements and Decay Constant Scaling}
\label{subsecbb}

To examine how the heavy-quark dynamics of the longitudinal wave
function enters the model prediction for the pseudoscalar decay
constant $f_P$, we consider the local axial-vector current matrix
element at zero transverse separation ($\mathbf{b}_\perp = 0$ or
$\zeta = 0$) within the light-front holographic framework:
\begin{equation}
f_P = \sqrt{\frac{2N_c}{\pi}} \left. \int_0^1 \, dx \, \Psi(x,
\mathbf{b}_\perp) \right|_{\mathbf{b}_\perp=0}, \label{eq401b}
\end{equation}
where $N_c=3$ is the number of colors. Factorization of the ground
state implies the boundary condition $\Psi(x, \zeta=0) =
(\kappa/\sqrt{\pi})X(x)$.

Specializing to a heavy--light pseudoscalar meson $B = (q\bar{b})$,
Eq. (\ref{eq401b}) simplifies directly to
\begin{equation}
f_B = \frac{\kappa\sqrt{2N_c}}{\pi} \int_0^1 \, dx \, X(x).
\label{eq402b}
\end{equation}
In contrast to prescriptions that incorporate explicit
spin-improvement terms, the current matrix element in this minimal
prescription involves only the longitudinal profile $X(x)$, without
additional inverse-momentum-fraction weights in the integrand. The
resulting heavy-quark asymptotic behaviors of the IMA and LMZV wave
functions can then be obtained from their respective normalization
conditions.

In the IMA ansatz, the longitudinal profile contracts into a narrow
peak centered at the saddle point $x_0 \sim m_q/m_B \sim m_B^{-1}$,
with an effective width shrinking as $\Delta x = \Delta\omega/m_B
\propto m_B^{-3/2}$.

The relativistic normalization condition
\begin{equation}
\int_0^1 \frac{\, dx}{x(1-x)}|X(x)|^2 = 1 \label{eq403b}
\end{equation}
determines the peak amplitude. To make the saddle-point scaling
explicit, we define the effective normalization factor
\begin{equation}
\widetilde{N} \equiv N \exp\left[ -\frac{(m_q+m_b)^2}{2\kappa^2}
\right], \label{eq404b}
\end{equation}
where the exponential factor is the value of the IMA exponential at
the saddle point. The normalization condition then gives
\begin{equation}
\widetilde{N} \propto (\Delta x)^{-1/2} \propto m_B^{3/4}, \qquad
X(x_0) \propto \widetilde{N}\sqrt{x_0(1-x_0)} \propto m_B^{1/4}.
\label{eq405b}
\end{equation}
Applying the saddle-point estimate $\int_0^1 \, dx \, X(x) \sim
X(x_0)\Delta x$, the integral scales as
\begin{equation}
\int_0^1 \, dx \, X(x) \sim X(x_0)\Delta x \sim m_B^{1/4} \cdot
m_B^{-3/2} \propto m_B^{-5/4}. \label{eq406b}
\end{equation}
Consequently, assuming a fixed confinement scale $\kappa$, the
asymptotic scaling of the decay constant in the IMA framework is
\begin{equation}
f_B \propto m_B^{-5/4}. \label{eq407b}
\end{equation}
Within this saddle-point estimate, this fractional power follows
from the narrowing longitudinal profile and the relativistic
normalization condition.

For the dynamical ground-state wave function $X_0(x) = N_0
x^{(\alpha_0+1)/2} (1-x)^{(\beta_0+1)/2}$, the integral can be
computed analytically in terms of the Euler beta function:
\begin{equation}
f_B = \frac{\kappa\sqrt{2N_c}}{\pi} N_0 \, B\left(
\frac{\alpha_0+3}{2}, \frac{\beta_0+3}{2} \right). \label{eq408b}
\end{equation}
In the static limit ($\beta_0 \sim 2m_b/ \sigma \to \infty$ at fixed
$\alpha_0 = 2m_q/\sigma$), the normalization constant scales as
\begin{equation}
N_0 \simeq
\frac{\beta_0^{(\alpha_0+1)/2}}{\sqrt{\Gamma(\alpha_0+1)}}.
\label{eq409b}
\end{equation}
Using the asymptotic property of the beta function for large second
arguments, $B(a,b) \simeq \Gamma(a)b^{-a}$, the matrix element
scales as follows:
\begin{equation}
\begin{aligned}
f_B &\propto \kappa N_0 \, B\left( \frac{\alpha_0+3}{2},
\frac{\beta_0+3}{2} \right) \sim \kappa \, \beta_0^{(\alpha_0+1)/2}
\, \beta_0^{-(\alpha_0+3)/2} \sim \kappa\beta_0^{-1} \sim
\mathcal{O}\left( \frac{\kappa\sigma}{m_B} \right).
\end{aligned}
\label{eq410b}
\end{equation}
This yields the heavy-quark scaling behavior in the LMZV
formulation:
\begin{equation}
f_B \propto m_B^{-1}. \label{eq411b}
\end{equation}

Comparing the two frameworks shows that neither reproduces the
canonical HQET scaling $f_B \propto m_B^{-1/2}$ when the soft
confinement scales are held fixed and the bare current matrix
element at zero impact separation is used. The origin of the
resulting powers differs between the two approaches: the LMZV wave
function yields an integer power law ($m_B^{-1}$), whereas the IMA
yields a fractional power ($m_B^{-5/4}$) associated with its
narrowing profile. These results illustrate that the heavy-quark
scaling of the decay constant is intimately tied to the longitudinal
structure of the LFWF and to the current prescription.

The scaling results obtained here should therefore be interpreted as
diagnostics of the bare light-front current within the two model
constructions, rather than as physical predictions for the decay
constant in the fully matched theory. A systematic matching to the
appropriate effective-theory current is required before drawing
conclusions about the physical heavy-quark scaling of $f_B$.

\section{Conclusions and Outlook}
\label{sec5}

The heavy-quark asymptotic behavior of light-front holographic wave
functions for heavy--light mesons has been examined with particular
emphasis on the leading-twist $B$-meson LCDA $\phi_B^+(\omega)$ and
the pseudoscalar decay constant $f_B$. The analysis has focused on
two different prescriptions for incorporating finite quark masses
into the holographic framework, namely the IMA and the longitudinal
variational construction based on the LMZV confining potential. A
central result is that the two approaches exhibit qualitatively
different behavior in the heavy-quark limit, even though they share
the same transverse holographic structure. In the IMA, the spectator
momentum remains localized at a soft scale, $\omega_0\to m_q$, but
the characteristic width of the distribution decreases as
$\Delta\omega\sim\sqrt{m_q\kappa^2/m_B}\propto m_B^{-1/2}$ at fixed
$\kappa$. Consequently, the longitudinal distribution becomes
increasingly narrow and approaches the limiting form
$\delta(\omega-m_q)$ as $m_b\to\infty$. The corresponding inverse
moment therefore approaches the finite value
$\lambda_{B,\mathrm{IMA}}\to m_q$, rather than vanishing. Thus, the
limitation identified in the IMA is associated with the collapse of
the longitudinal width rather than with a vanishing inverse moment.
Maintaining a finite width within this scaling estimate would
require an explicit heavy-quark dependence of the transverse
confinement scale, rather than a fixed soft scale
$\kappa\sim\Lambda_{\mathrm{QCD}}$.

A different behavior is obtained from the LMZV longitudinal
confining dynamics. In this case, the heavy-quark limit produces a
finite-width LP profile, $\mathcal{X}_{\mathrm{LP}}(\omega)$, whose
characteristic momentum scale remains of order $\sigma$. The peak
approaches $\omega_0\to m_q+\sigma/2$, while the limiting inverse
moment is determined by the parameters of the longitudinal
Hamiltonian:
\begin{equation}
\lambda_{B,\mathrm{LMZV}}^{\mathrm{LP}} = m_q + \frac{\sigma}{2} =
\mathcal{O}(\Lambda_{\mathrm{QCD}})\,. \label{eq501}
\end{equation}
Within this construction, the Gamma-type profile retains a
non-singular soft distribution as the heavy-quark mass increases,
without the width collapse found in the IMA. The relation between
the inverse moment and the model parameters provides a direct
characterization of the longitudinal dynamics encoded in the LMZV
wave function. This result should be understood as a
parameter-dependent prediction within the model, rather than as a
parameter-independent determination of $\lambda_B$.

The analysis of the decay constant leads to a separate conclusion.
The stability of the longitudinal wave function does not, by itself,
guarantee the correct heavy-quark scaling of a local current matrix
element. For the minimal light-front current evaluated at zero
impact separation ($\mathbf{b}_\perp = 0$), the matrix element
depends directly on the longitudinal overlap integral $\int_0^1 \,
dx\, X(x)$ without explicit spin-enhancement factors. Within this
current prescription, the two longitudinal constructions exhibit
different noncanonical powers: $f_B\propto m_B^{-5/4}$ for the IMA
and $f_B\propto m_B^{-1}$ for the LMZV framework. Thus, neither the
IMA nor the LMZV construction, when combined with the bare
light-front current and fixed soft confinement parameters,
reproduces the canonical HQET scaling $f_B\sim m_B^{-1/2}$. These
results demonstrate that the heavy-quark behavior of the
longitudinal distribution and that of the local current matrix
element constitute distinct consistency requirements.

The mismatch in the decay-constant scaling indicates that the bare
light-front current evaluated at zero transverse separation is not
sufficient to establish the physical heavy-quark-limit behavior of
$f_B$. A systematic treatment of the current and its matching to the
appropriate effective-theory operator is therefore needed before a
physical prediction for this scaling can be claimed. The present
analysis does not establish whether such matching modifies the
power-law behavior obtained from the bare current. Within the IMA,
introducing a heavy-quark dependence of the transverse confinement
scale $\kappa(m_b)$ would provide an alternative model prescription,
but would change the assumption of a fixed soft confinement scale.
For the LMZV framework, the finite LP profile and the
parameter-dependent result
$\lambda_{B,\mathrm{LMZV}}^{\mathrm{LP}}=m_q+\sigma/2$ provide a
useful starting point for a separate investigation of current
matching and scale dependence, without requiring an immediate
modification of the longitudinal construction.

Several directions follow from these findings. Formulating the
light-front current directly in terms of effective-theory currents
could clarify how heavy-quark spin symmetry and short-distance QCD
corrections enter the relativistic light-front description. A
systematic analysis of the matching between the light-front current
and the HQET current would also be needed to determine how the
current matrix element should be interpreted in the heavy-quark
limit and whether the appropriate HQET power counting ($f_B \propto
m_B^{-1/2}$) is recovered. On the phenomenological side, the LP LMZV
LCDA and its parameter-dependent inverse moment provide a starting
point for future studies of exclusive $B$-meson processes, including
$B\to\gamma\ell\nu$ and $B\to\pi\ell\nu$. Such applications could
test the sensitivity of observables to the LCDA shape and to the
longitudinal confinement parameters, while providing a setting in
which the model's nonperturbative input and the treatment of the
current can be examined together.

In summary, the longitudinal variational holographic construction
retains a finite-width soft distribution in the heavy-quark limit,
whereas the IMA distribution becomes parametrically narrow at fixed
soft transverse confinement scale. At the same time, the
decay-constant analysis shows that a stable longitudinal profile
alone is insufficient to reproduce the canonical HQET scaling of the
local current matrix element. Distinguishing these two aspects is
essential for assessing the heavy-quark consistency of the
holographic description and motivates a treatment in which the
nonperturbative wave-function dynamics and the matching of the
current to the effective theory are examined separately.


\begin{thebibliography}{99}

\bibitem{Buchalla1995}
G. Buchalla, A. J. Buras, and M. E. Lautenbacher, Rev. Mod. Phys.
\textbf{68}, 1125 (1996) [arXiv:hep-ph/9512380].

\bibitem{Artuso2009}
M. Artuso, E. Barberio, and S. Stone, PMC Phys. A \textbf{3}, 3
(2009) [arXiv:0902.3743 [hep-ph]].

\bibitem{Brod2014}
J. Brod, PoS \textbf{Modave2014}, 001 (2015) [arXiv:1504.05948].

\bibitem{Beneke1999}
M. Beneke, M. Buchalla, M. Neubert, and C. T. Sachrajda, Phys. Rev.
Lett. \textbf{83}, 1914 (1999) [arXiv:hep-ph/9905312].

\bibitem{Beneke2000}
M. Beneke and Th. Feldmann, Nucl. Phys. B \textbf{592}, 3 (2001)
[arXiv:hep-ph/0008255].

\bibitem{Bauer2000}
C. W. Bauer, S. Fleming, and M. E. Luke, Phys. Rev. D \textbf{63},
014006 (2000) [arXiv:hep-ph/0005275].

\bibitem{Bauer2001}
C. W. Bauer, D. Pirjol, and I. W. Stewart, Phys. Rev. D \textbf{65},
054022 (2002) [arXiv:hep-ph/0109045].

\bibitem{Beneke2002}
M. Beneke, A. P. Chapovsky, M. Diehl, and Th. Feldmann, Nucl. Phys.
B \textbf{643}, 431--476 (2002) [arXiv:hep-ph/0206152].

\bibitem{Grozin1996}
A. G. Grozin and M. Neubert, Phys. Rev. D \textbf{55}, 272 (1997)
[arXiv:hep-ph/9607366].

\bibitem{DescotesGenon2002}
S. Descotes-Genon and C. T. Sachrajda, Nucl. Phys. B \textbf{650},
356--390 (2003) [arXiv:hep-ph/0209216].

\bibitem{Korchemsky1999}
G. P. Korchemsky, D. Pirjol, and T. M. Yan, Phys. Rev. D
\textbf{61}, 114510 (2000) [arXiv:hep-ph/9911427].

\bibitem{Lunghi2002}
E. Lunghi, D. Pirjol, and D. Wyler, Nucl. Phys. B \textbf{649}, 349
(2003) [arXiv:hep-ph/0210091].

\bibitem{Bosch2003}
S. W. Bosch, R. J. Hill, B. O. Lange, and M. Neubert, Phys. Rev. D
\textbf{67}, 094014 (2003) [arXiv:hep-ph/0301123].

\bibitem{Beneke2011}
M. Beneke and J. Rohrwild, Eur. Phys. J. C \textbf{71}, 1818 (2011)
[arXiv:1110.3228 [hep-ph]].

\bibitem{Wang2018}
Y. M. Wang and Y. L. Shen, JHEP \textbf{05}, 184 (2018)
[arXiv:1803.06667 [hep-ph]].

\bibitem{Braun2003}
V. M. Braun, D. Y. Ivanov, and G. P. Korchemsky, Phys. Rev. D
\textbf{69}, 034014 (2004) [arXiv:hep-ph/0309330].

\bibitem{Khodjamirian2006}
A. Khodjamirian, T. Mannel, and N. Offen, Phys. Rev. D \textbf{75},
054013 (2007) [arXiv:hep-ph/0611193].

\bibitem{Gubernari2018}
N. Gubernari, A. Kokulu, and D. van Dyk, JHEP \textbf{01}, 150
(2019) [arXiv:1811.00983 [hep-ph]].

\bibitem{Kawamura2018}
H. Kawamura, J. Kodaira, C. F. Qiao, and K. Tanaka, Phys. Lett. B
\textbf{523}, 111 (2001) [arXiv:hep-ph/0109181]; Erratum: Phys.
Lett. B \textbf{536}, 344 (2002).

\bibitem{Wang2019}
W. Wang, Y. M. Wang, J. Xu, and S. Zhao, Phys. Rev. D \textbf{102},
011502(R) (2020) [arXiv:1908.09933 [hep-ph]].

\bibitem{Brodsky2006}
S. J. Brodsky and G. F. de T'eramond, Phys. Rev. Lett. \textbf{96},
201601 (2006) [arXiv:hep-ph/0602252].

\bibitem{DeTeramond2008}
G. F. de T'eramond and S. J. Brodsky, Phys. Rev. Lett. \textbf{102},
081601 (2009) [arXiv:0809.4899 [hep-ph]].

\bibitem{Brodsky2014}
S. J. Brodsky, G. F. de T'eramond, H. G. Dosch, and J. Erlich, Phys.
Rep. \textbf{584}, 1 (2015) [arXiv:1407.8131 [hep-ph]].

\bibitem{Karch2006}
A. Karch, E. Katz, D. T. Son, and M. A. Stephanov, Phys. Rev. D
\textbf{74}, 015005 (2006) [arXiv:hep-ph/0602229].

\bibitem{Brodsky2008}
S. J. Brodsky and G. F. de T'eramond, Phys. Rev. D \textbf{77},
056007 (2008) [arXiv:0707.3859 [hep-ph]].

\bibitem{Brodsky2008a}
S. J. Brodsky and G. F. de T'eramond, Subnucl. Ser. \textbf{45}, 139
(2009) [arXiv:0802.0514 [hep-ph]].

\bibitem{Ahmady2018}
M. Ahmady, C. Mondal, and R. Sandapen, Phys. Rev. D \textbf{98},
034010 (2018) [arXiv:1805.08911 [hep-ph]].

\bibitem{Ahmady2020}
M. Ahmady, S. Lord, and R. Sandapen, Phys. Rev. D \textbf{101},
034013 (2020) [arXiv:1911.12450 [hep-ph]].

\bibitem{Isgur1989}
N. Isgur and M. B. Wise, Phys. Lett. B \textbf{232}, 113 (1989).

\bibitem{Neubert1994}
M. Neubert, Phys. Rep. \textbf{245}, 259 (1994)
[arXiv:hep-ph/9306320].

\bibitem{Li2016}
Y. Li, P. Maris, X. Zhao, and J. P. Vary, Phys. Lett. B
\textbf{758}, 118 (2016) [arXiv:1509.07212 [hep-ph]].

\bibitem{Jia2019}
S. Jia and J. P. Vary, Phys. Rev. C \textbf{99}, 035206 (2019)
[arXiv:1811.08512 [nucl-th]].

\bibitem{Brodsky1989}
S. J. Brodsky and G. P. Lepage, in \textit{Perturbative Quantum
Chromodynamics}, edited by A. H. Mueller (World Scientific,
Singapore, 1989), pp. 93--240.

\bibitem{Lepage1980}
G. P. Lepage and S. J. Brodsky, Phys. Rev. D \textbf{22}, 2157
(1980).

\bibitem{Li2022}
Y. Li and J. P. Vary, Phys. Lett. B \textbf{825}, 136860 (2022)
[arXiv:2103.09993 [hep-ph]].

\bibitem{DeTeramond2021}
G. F. de T'eramond and S. J. Brodsky, Phys. Rev. D \textbf{104},
116009 (2021) [arXiv:2103.10950 [hep-ph]].

\bibitem{Brodsky2016}
S. J. Brodsky, G. F. de T\'eramond, H. G. Dosch and C. Lorc\'e,
Phys. Lett. B \textbf{759}, 171 (2016) [arXiv:1604.06746 [hep-ph]].

\bibitem{DeTeramond2014}
G. F. de T\'eramond, H. G. Dosch and S. J. Brodsky, Phys. Rev. D
\textbf{91}, 045040 (2015) [arXiv:1411.5243 [hep-ph]].

\bibitem{Chang2017}
Q. Chang, S. Xu and L. Chen, Nucl. Phys. B \textbf{921}, 454 (2017)
[arXiv:1805.02011 [hep-ph]].

\bibitem{Chang2017a}
Q. Chang, S. J. Brodsky and X. Q. Li, Phys. Rev. D \textbf{95},
094025 (2017) [arXiv:1612.05298 [hep-ph]].

\bibitem{Nielsen2018}
M. Nielsen, S. J. Brodsky, G. F. de T\'eramond, H. G. Dosch, F. S.
Navarra and L. Zou, Phys. Rev. D \textbf{98}, 034002 (2018)
[arXiv:1805.11567 [hep-ph]].

\bibitem{Ahmady2021}
M. Ahmady, S. Kaur, S. L. MacKay, C. Mondal and R. Sandapen, Phys.
Rev. D \textbf{104}, 074013 (2021) [arXiv:2108.03482 [hep-ph]].

\bibitem{Cheung1996}
C.-Y. Cheung, C.-W. Hwang and W.-M. Zhang, Z. Phys. C \textbf{75},
657 (1997) [arXiv:hep-ph/9602309].

\end{thebibliography}
\end{document}